\documentclass[aps,pra,reprint,showpacs,notitlepage,superscriptaddress,twocolumn]{revtex4-2}
\usepackage{amssymb}
\usepackage{mathrsfs}
\usepackage{amsfonts}
\usepackage{graphicx}
\usepackage{amsmath}
\usepackage{dcolumn}
\usepackage{color}
\usepackage{subfigure}
\usepackage{epsfig}
\usepackage{soul}
\usepackage[colorlinks,linkcolor=blue,citecolor=blue,hyperindex,bookmarks=false,pdfstartview=FitH]{hyperref}

\providecommand{\U}[1]{\protect\rule{.1in}{.1in}}

\newcommand{\figpanel}[2]{\hyperref[#1]{\ref*{#1}(#2)}}
\renewcommand{\eqref}[1]{\hyperref[#1]{(\ref*{#1})}}

\begin{document}

\title{Controlling entanglement by phase engineering in giant-atom waveguide}

\author{Peng-Fei Wang}
\affiliation{College of Physics and Electronic Engineering, Hainan Normal University, Haikou 571158, People’s Republic of China}
\author{Lei Huang}
\affiliation{College of Physics and Electronic Engineering, Hainan Normal University, Haikou 571158, People’s Republic of China}
\author{Miao-Miao Wei}
\affiliation{College of Physics and Electronic Engineering, Hainan Normal University, Haikou 571158, People’s Republic of China}
\author{Hong Yang}
\affiliation{College of Physics and Electronic Engineering, Hainan Normal University, Haikou 571158, People’s Republic of China}
\author{Dong Yan}
\email{yand@hainnu.edu.cn}
\affiliation{College of Physics and Electronic Engineering, Hainan Normal University, Haikou 571158, People’s Republic of China}

\date{\today}

\begin{abstract}

We investigate the entanglement dynamics of two giant atoms coupled to a common waveguide. By introducing additional phase modulation at each coupling point, every photon propagation path is jointly controlled by two distinct coupling phases, enabling precise and flexible manipulation of the entanglement evolution. This phase engineering induces destructive interference among different paths, leading to entanglement dynamics in nested giant atoms that become equivalent to those of small atoms, as well as dynamical equivalence between separated and braided configurations. Furthermore, the proposed scheme significantly enhances the robustness of entanglement against variations in the phase shift, offering a practical route to generate stable entanglement and enabling quantum devices with programmable propagation and controllable memory effects.  
\end{abstract}

\maketitle
\section{Introduction}
\label{section1}
Quantum entanglement is a fundamental nonclassical correlation in quantum mechanics~\cite{PhysRev.47.777,RevModPhys.81.865,duarte2021quantum} and serves as a key resource for quantum communication~\cite{bennett1993teleporting}, quantum computation~\cite{nielsen2010quantum}, and quantum networks~\cite{kimble2008quantum}. Extensive efforts have been devoted to generating and manipulating entanglement across various physical platforms, including cavity quantum electrodynamics (QED)~\cite{raimond2001manipulating,matsukevich2005entanglement,volz2006observation,wilk2007single}, trapped ion~\cite{leibfried2003quantum,blinov2004observation}, quantum dot~\cite{gao2012observation,lodahl2015interfacing}, and superconducting qubits~\cite{you2011atomic,xiang2013hybrid,gu2017microwave}. Waveguide QED, in particular, provides an ideal platform for mediating long-range interactions between quantum emitters via one-dimensional photonic modes,  offering strong light–matter coupling and the potential for scalable quantum networks~\cite{roy2017colloquium,chang2018colloquium,PhysRevLett.106.020501}. 

In parallel, the advent of artificial atoms with extended coupling points, termed giant atoms, has opened new avenues in waveguide QED~\cite{fiveyear}. In these systems, the breakdown of the dipole approximation and the presence of multiple coupling points lead to rich quantum interference phenomena, such as frequency-dependent relaxation~\cite{fiveyear,gustafsson2014propagating,LambAFK}, decoherence-free interactions~\cite{braidedkannan2020waveguide,NoriGA,FCdeco,complexDFI,PhysRevA.105.023712,PhysRevA.107.013710,PhysRevResearch.6.043222,leonforte2024quantum}, chiral spontaneous emission~\cite{PhysRevX.13.021039,chen2023giant,du2022giant,wang2022chiral,wang2024unconventional,crzs-k718}, anomalous single-photon scattering~\cite{ZhaoWGA,DLlambda1,DLlambda2,JiaGA1,YinGA,ZhuYTGA,CYTcp,ZhouJGA,PhysRevA.109.063703,nnmt-8492}, non-Markovian dynamics~\cite{andersson2019non,guo2017giant,du2022giantPRR,xu2024catch,PhysRevA.109.023712}, and unconventional bound states~\cite{WXchiral1,oscillating1,PhysRevA.107.023716,PhysRevA.111.053710}. Giant atoms have been experimentally realized in superconducting circuits coupled to surface acoustic waves~\cite{gustafsson2014propagating,andersson2019non} and microwave transmission lines~\cite{braidedkannan2020waveguide,PhysRevX.13.021039,vadiraj2021engineering}, and have been proposed platforms including optical lattices~\cite{gonzalez2019engineering}, coupled waveguide arrays~\cite{longhi2020photonic,PhysRevA.107.023716}, Rydberg atoms~\cite{chen2023giant,chen2024giant,cc46-f919}, synthetic photonic dimensions~\cite{du2022giant,xiao2022bound}, and spin ensembles~\cite{wang2022giant}. These features establish giant atoms a promising platform for entanglement engineering and quantum networks applications~\cite{PhysRevA.108.023728,PhysRevA.106.063703,cai2023nonreciprocal,PhysRevLett.130.053601}.

Previous studies have shown that entanglement properties in giant-atom systems depend strongly on their topological configuration~\cite{PhysRevA.108.023728}. Non-Markovian effects often induce phenomena such as sudden death and revival of entanglement at early times, followed by convergence to a steady state~\cite{PhysRevA.106.063703}. Chiral couplings can further lead to nonreciprocal~\cite{cai2023nonreciprocal} or enhanced entanglement~\cite{liu2024entanglement}. In single-emitter systems, chirality is typically characterized by the phase difference between two coupling points~\cite{roccati2024controlling}, and this description has been extended to multi-atom setups~\cite{cai2023nonreciprocal}. However, in multi-atom systems, each propagation path between atoms is simultaneously influenced by two independent coupling phases. Describing chirality solely via their difference is insufficient to capture the full role of phase modulation at each point.

In this work, following the approach of Ref.~\cite{PhysRevX.13.021039}, we treat the coupling phases at each atom–waveguide interaction point as independent control knobs. The resulting atomic dynamics reveal that each propagation path is governed by a combination of the waveguide phase shift, time delay, and two coupling phases. By appropriately tuning these phases, destructive interference between different paths can be engineered, suppressing non-Markovian effects and stabilizing entanglement. This leads to reduced early-time oscillations and faster convergence to steady-state entanglement. Moreover, with suitable phase engineering, the nested configuration becomes completely immune to certain waveguide parameters, and the entanglement dynamics of two giant atoms becomes equivalent to that of two small atoms. In the separated configuration, the dynamics exhibit robustness against phase shift variations, and different topological configurations can yield identical entanglement behavior under proper coupling phase conditions.

This paper is organized as follows. In Sec.~\ref{section2}, we introduce the system model and Hamiltonian, derive the dynamical equations for the two giant atoms, and present the expression for the concurrence. Sec.~\ref{section3} focuses on the nested configuration, where destructive interference induced by coupling phases suppresses non-Markovian effects and yields entanglement dynamics equivalent to those of small atoms. In Sec.~\ref{section4}, we examine the robustness of entanglement against specific waveguide parameters in the nested configuration and against the phase shift in the separated configuration. Sec.~\ref{section5} demonstrates that identical entanglement dynamics can be realized in separate and braided configurations via appropriate coupling phase modulation. Finally, we present a brief discussion and conclusion in Sec.~\ref{section6}.

\section{MODEL AND DYNAMICAL EQUATIONS}
\label{section2}
\begin{figure}[htbp]
    \centering
\includegraphics[width=0.85\linewidth]{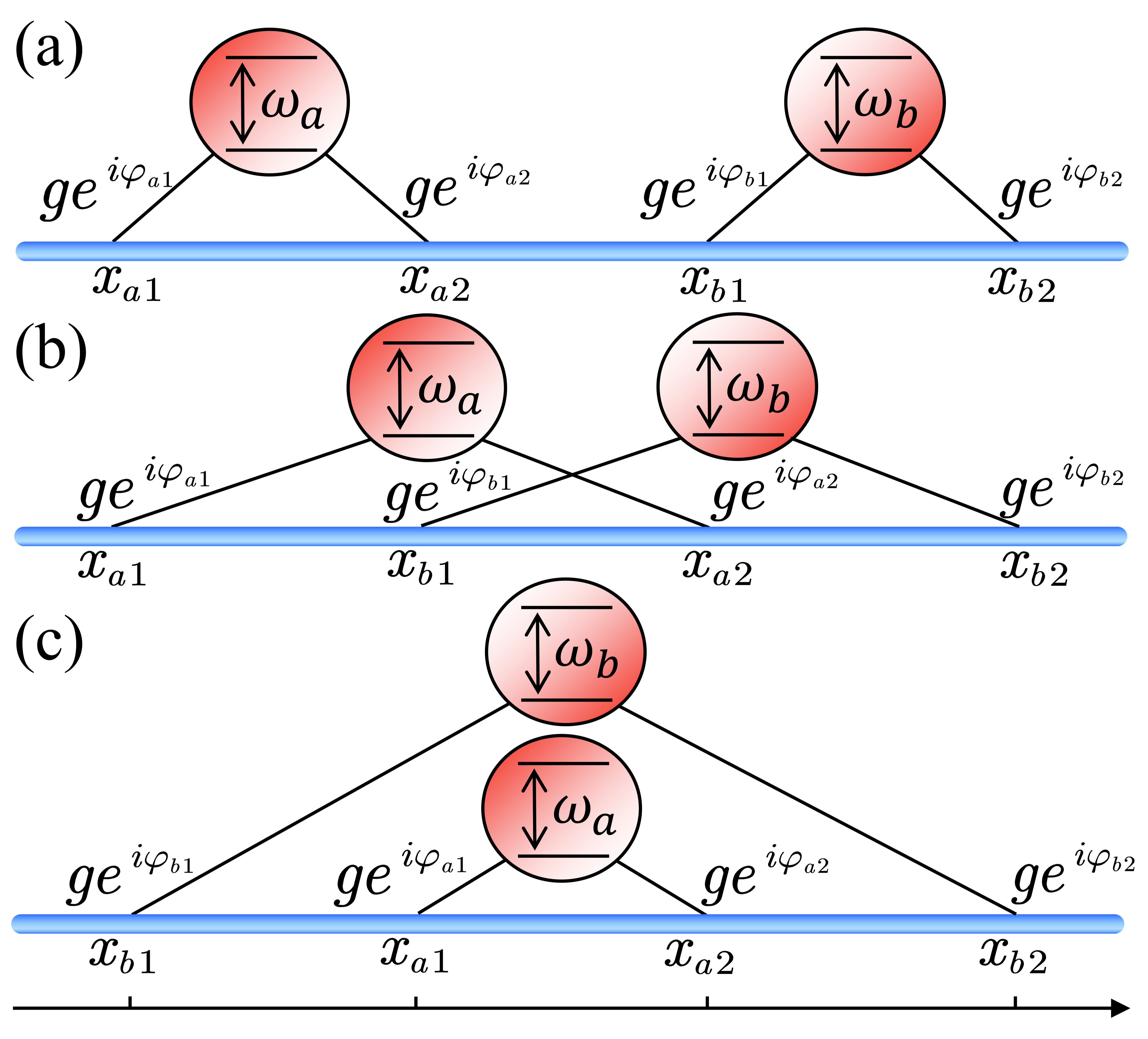}
    \caption{Three typical configurations of two two-level giant atoms with transition frequencies $\omega_a$ and $\omega_b$ coupled to a common waveguide. (a) Separate coupling, (b) Braided coupling, and (c) Nested coupling. Each giant atom interacts with the waveguide via two coupling points labeled $x_{jn}$, where $j = a, b$ denotes the atom and $n = 1,2$ indexes its coupling points. The atom–waveguide coupling coefficient at position $x_{jn}$ is $g e^{i\varphi_{jn}}$, with $\varphi_{jn}$ the coupling phase.}
    \label{FIG1}
\end{figure}
We consider two giant atoms coupled to a shared waveguide. Each atom interacts with the waveguide at two spatially separated points $x_{jn}$ ($j=a,b$; $n=1,2$. As demonstrated theoretically~\cite{roccati2024controlling} and experimentally~\cite{PhysRevX.13.021039}, a coupling phase $\varphi_{jn}$ can be imprinted at each interaction points. Depending on the spatial arrangement of these coupling points along the waveguide, three distinct configurations emerge: separate, braided, and nested, as illustrated in Fig.~\ref{FIG1}. 

The Hamiltonian of the system ($\hbar = 1$ hereafter) is 
\begin{equation}
\begin{split}
   H=&\sum_{j=a,b}{\omega _j\hat{\sigma}_{j}^{+}\hat{\sigma}_{j}^{-}}+\sum_k{\omega_k\hat{c}_{k}^{\dagger}\hat{c}_k}
   \\
&+g\sum_{j=a,b}{\sum_{n=1,2}{\sum_k{\left[ e^{i\left( \varphi _{jn}+kx_{jn} \right)}\hat{\sigma}_{j}^{+}\hat{c}_k+\mathrm{H}.\mathrm{c}. \right]}}}.
\end{split}
\label{EQ1}
\end{equation}
The first two terms describe the free Hamiltonians of the giant atoms and the waveguide field, respectively, while the third term accounts for their mutual interaction. Here, $\omega_j$ is the transition frequency between the excited state $|e\rangle_j$ and the ground state $|g\rangle_j$ of atom $j$. The operators $\sigma_j^- = |g\rangle_{jj}\langle e|$ ($\sigma_j^+ = |e\rangle_{jj}\langle g|$) are the lowering (raising) operators, and $c_k$ ($c_k^\dagger$) annihilates (creates) a waveguide photon with wave vector $k$ and frequency $\omega_k$. The real coupling coefficient $g$ characterizes the atom–waveguide interaction strength.

Given the conservation of total excitation number, we restrict our analysis to the single-excitation subspace. The system state can then be written as

\begin{equation}
\begin{split}
|\psi \left( t \right) \rangle =\left[ \sum_{j=a,b}{c_j(t)\hat{\sigma}_{j}^{+}}e^{-i\omega _jt}+\sum_k{u_k(t)\hat{c}_{k}^{\dagger}}e^{-i\omega _kt} \right] |G\rangle,
\end{split}
\label{EQ2}
\end{equation}
where $c_j(t)$ is the excitation amplitude of atom $j$, $u_k(t)$ is the amplitude of the waveguide mode with wavevector $k$, and $|G\rangle$ denotes the ground state where both atoms are in their ground states and the waveguide field is in the vacuum state. Solving the Schr\"odinger equation yields

\begin{equation}
\begin{aligned}
\dot{c}_j(t)&=-ig\sum_{n=1,2}{\sum_k{e^{i\left( \varphi_{jn}+kx_{jn} \right)}e^{-i\left( \omega _k-\omega _j \right) t}u_k\left( t \right)}},\\
\dot{u}_k(t)&=-ig\sum_{n=1,2}{\sum_{j=a,b}{e^{-i(\varphi_{jn}+kx_{jn})}e^{i(\omega _k-\omega _j)t}c_j\left( t \right)}}.
\end{aligned}
\label{EQ3}
\end{equation}
We consider an initially excited atoms, so $u_k(0)=0$. Integrating Eq.~\eqref{EQ3} gives the formal solution for $u_k(t)$.
\begin{equation}
\begin{split}
\begin{aligned}
u_k(t)=-ig\sum_{n=1,2}{\sum_{j=a,b}{e^{-i\left( \varphi _{jn}+kx_{jn} \right)}}}\\
\times \int_0^t{e^{-i(\omega _k-\omega _j)(t-s)}c_j(s)ds}.
\end{aligned}
\end{split}
\label{EQ4}
\end{equation}

Following the Weisskopf–Wigner approximation, we linearize the dispersion relation around $\omega_0$ as $\omega _k\approx \omega _0+\nu =\omega _0+(k-k_0)v_g$, where $k_0$ is the wavevector corresponding to $\omega_0$ and $v_g$ is the group velocity~\cite{PhysRevLett.95.213001,PhysRevA.79.023837,agarwal2012quantum}. Assuming $\omega_j \equiv \omega_0$ and substituting the formal solution of $u_k(t)$ into Eq.~\eqref{EQ3}, we obtain delay differential equations for the atomic amplitudes:

\begin{widetext}
\begin{equation}
\begin{aligned}
\dot{c}_a(t) = {} & -\Gamma c_a(t)
-\Gamma \cos(\varphi_{a2}-\varphi_{a1}) e^{i\theta_{a1,a2}}
c_a\bigl(t-\tau_{a1,a2}\bigr)\Theta\bigl(t-\tau_{a1,a2}\bigr) \\
& -\frac{\Gamma}{2}\sum_{m,n=1}^{2}
e^{i[(\varphi_{an}-\varphi_{bm})+\theta_{an,bm}]}
c_b\bigl(t-\tau_{an,bm}\bigr)\Theta\bigl(t-\tau_{an,bm}\bigr), \\
\dot{c}_b(t) = {} & -\Gamma c_b(t)
-\Gamma \cos(\varphi_{b2}-\varphi_{b1}) e^{i\theta_{b1,b2}}
c_b\bigl(t-\tau_{b1,b2}\bigr)\Theta\bigl(t-\tau_{b1,b2}\bigr) \\
& -\frac{\Gamma}{2}\sum_{m,n=1}^{2}
e^{i[(\varphi_{bm}-\varphi_{an})+\theta_{an,bm}]}
c_a\bigl(t-\tau_{an,bm}\bigr)\Theta\bigl(t-\tau_{an,bm}\bigr).
\end{aligned}
\label{EQ5}
\end{equation}
\end{widetext}
Here, $\Gamma = 4\pi g^2/v_g$ is the spontaneous emission rate, and $\Theta$ is the Heaviside step function. We neglect dissipation into non-waveguide modes and assume atomic lifetime much longer than photon-exchange timescale. The time delays are defined as $\tau_{j1,j2}=|x_{j1}-x_{j2}|/v_g$ (within the same atom) and $\tau_{an,bm}=|x_{an}-x_{bm}|/v_g$ (between different atoms). The corresponding phase shifts are $\theta_{j1,j2}=k_0|x_{j1}-x_{j2}|$ and $\theta_{an,bm}=k_0|x_{an}-x_{bm}|$. In Eq.~\eqref{EQ5}, the first term describes spontaneous emission. The second term accounts for photon emission and reabsorption between coupling points of the same atom, while the remaining terms represent photon-mediated interactions between the two atoms. All processes except spontaneous emission depend on the coupling phases $\varphi_{jn}$.

By tracing out the waveguide modes, the reduced density matrix of the two atoms can be written in the basis ${|e\rangle_a|e\rangle_b, |e\rangle_a|g\rangle_b, |g\rangle_a|e\rangle_b, |g\rangle_a|g\rangle_b}$ as:

\begin{equation}
\begin{split}
\begin{aligned}
\hat{\rho}(t)=\left( \begin{matrix}
	0&		0&		0&		0\\
	0&		|c_a(t)|^2&		c_a(t)c_{b}^{*}(t)&		0\\
	0&		c_{a}^{*}(t)c_b(t)&		|c_b(t)|^2&		0\\
	0&		0&		0&		1-|c_a(t)|^2-|c_b(t)|^2\\
\end{matrix} \right).
\end{aligned}
\end{split}
\label{EQ6}
\end{equation}
Entanglement between the two atoms is quantified by the concurrence~\cite{PhysRevLett.130.053601,PhysRevLett.80.2245,PhysRevA.63.052302}:
\begin{equation}
\begin{split}
\begin{aligned}
C(t) &= \max\{0, \sqrt{\lambda_1} - \sqrt{\lambda_2} - \sqrt{\lambda_3} - \sqrt{\lambda_4} \}\\
&=2|c_a(t)c_{b}^{*}(t)|,
\end{aligned}
\end{split}
\label{EQ7}
\end{equation}
where $\lambda_i$ are the eigenvalues of $\sqrt{\rho} (\sigma_y \otimes \sigma_y) \rho^* (\sigma_y \otimes \sigma_y)\sqrt{\rho}$, with $\sigma_y$ the Pauli matrix. Equation~\eqref{EQ7} follows from the X-type structure of the density matrix~\cite{yu2007evolution}.

Thus, entanglement dynamics are governed by the atomic excitation amplitudes. Combined with Eq.~\eqref{EQ5}, we can systematically explore the effects of phase shift $\theta$, time delay $\Gamma\tau$, and coupling phase $\varphi_{jn}$ on entanglement.  In particular, the coupling phases can selectively activate or suppress propagation paths, leading to richer and more controllable entanglement behavior.

\section{Phase-Controlled Non-Markovian Entanglement Dynamics in Nested Giant Atoms}
\label{section3}

\begin{figure}[htbp]
    \centering
    \includegraphics[width=0.85\linewidth]{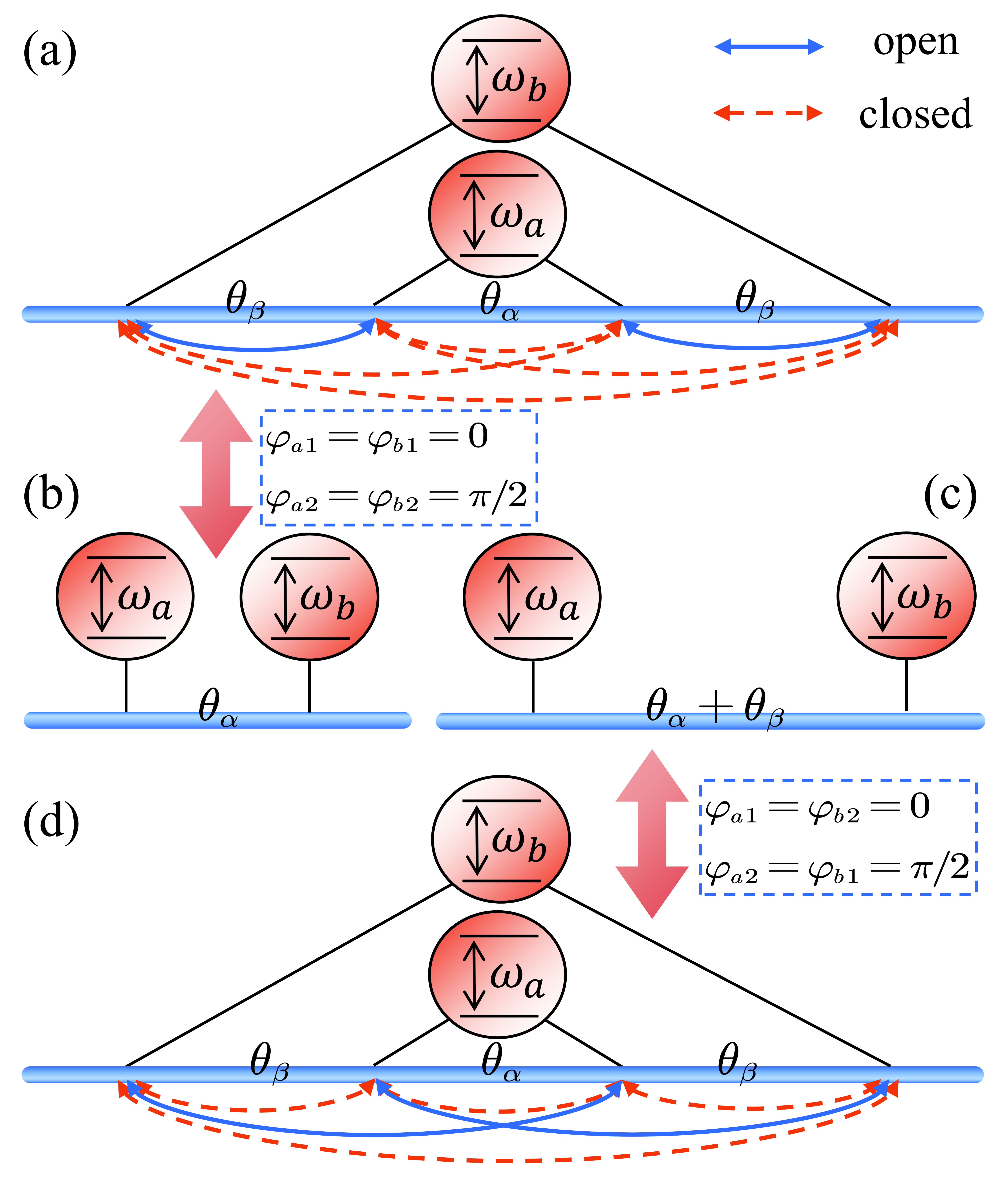}
    \caption{(a) Nested giant-atom configuration. For specific coupling phases, only a few propagation paths remain active (blue solid lines), while most are suppressed by destructive interference (red dashed lines). (b) Entanglement dynamics of two small atoms coupled to a waveguide with phase shift $\theta_{\alpha}$ between them, identical to the corresponding case in (a). (c) Entanglement dynamics of two small atoms with phase shift $\theta_{\alpha}+\theta_{\beta}$, matching the case in (d). (d) Another set of coupling-phase conditions leading to a different pattern of active and suppressed paths.}
    \label{FIG2}
\end{figure}

We first define the relevant propagation parameters for the nested configuration [see Fig.~\figpanel{FIG2}{a}]. For atom $a$, the phase shift between its two coupling points is denoted by $\theta_{\alpha}$, with corresponding time delay $\Gamma\tau_{\alpha}$. For paths connecting atoms $a$ and $b$, the phase shift between adjacent coupling points is $\theta_{\beta}$, and the associated time delay is $\Gamma\tau_{\beta}$.

In the conventional giant-atom equations, each propagation path contributes independently, governed solely by the phase shift and time delay. Introducing coupling phases adds an extra modulation to each path. When these modulations are appropriately chosen, destructive interference can render certain paths inactive. Even suppressing a few paths can significantly alter the non-Markovian dynamics, as we now show.

\begin{figure}[htbp]
    \centering
    \includegraphics[width=1\linewidth]{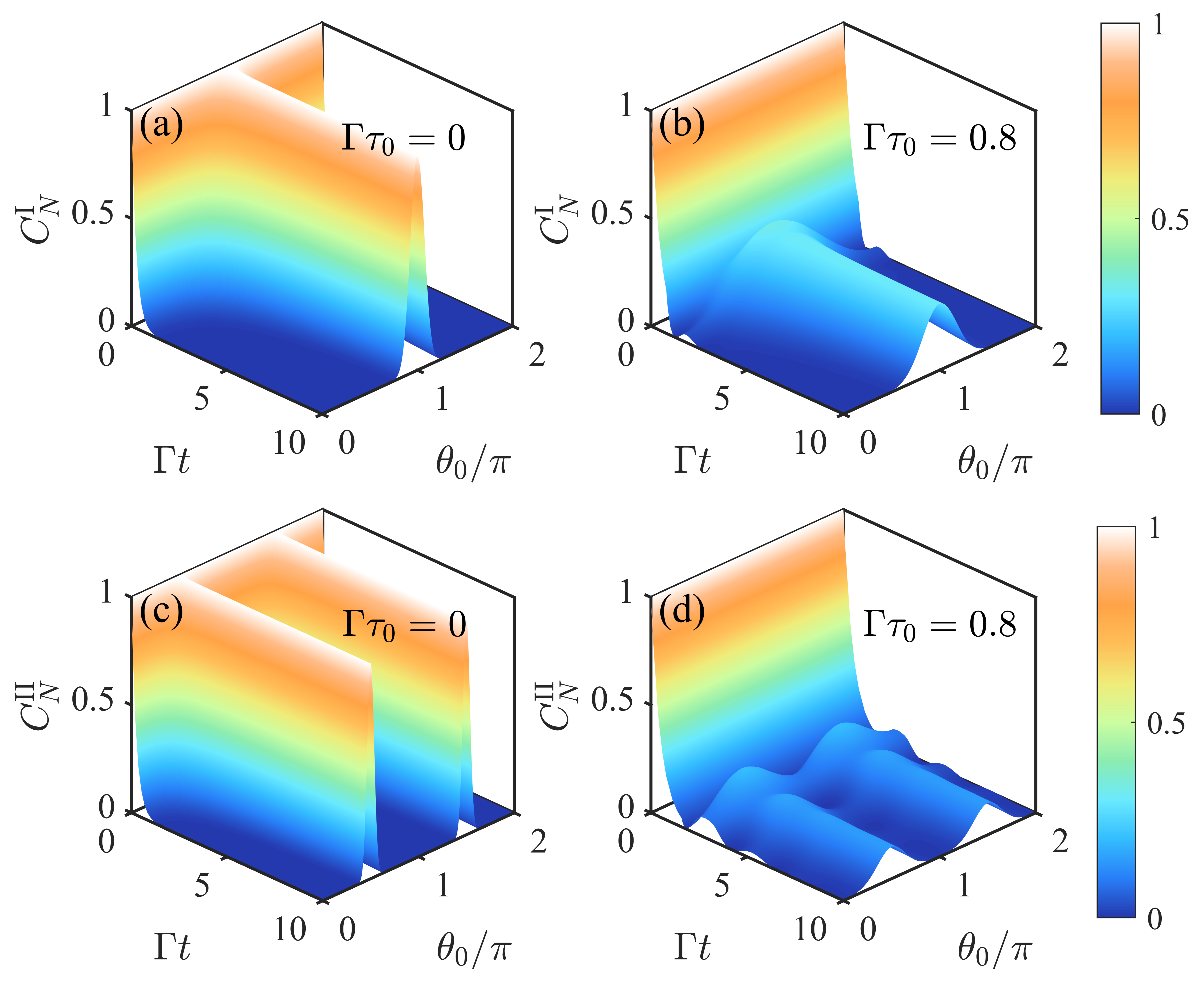}
    \caption{Concurrence $C_N^{\mathrm{I}}$ ($C_N^{\mathrm{II}}$) as a function of the evolution time $\Gamma t$ and the phase shift $\theta_0$ for different coupling phases and time delays $\Gamma \tau_0$. (a) and (c): $\Gamma\tau_0=0$; (b) and (d): $\Gamma\tau_0=0.8$. Case I: $\varphi _{a1}=\varphi _{b1}=0$, $\varphi _{a2}=\varphi _{b2}=\pi/2$. Case II: $\varphi _{a1}=\varphi _{b2}=0$, $\varphi _{a2}=\varphi _{b1}=\pi/2$. Initial state: $|+\rangle = (|eg\rangle + |ge\rangle)/\sqrt{2}$.}
    \label{FIG3}
\end{figure}

To simplify the analysis, we set $\theta_{\alpha}=\theta_{\beta}=\theta_{0}$ and $\tau_{\alpha}=\tau_{\beta}=\tau_{0}$ in this section. For Case I, the additional phase-modulation parameters are chosen as $\varphi_{a1}=\varphi_{b1} = 0$ and $\varphi_{a2}=\varphi_{b2} = \pi/2$. Under these conditions, Eq.~\eqref{EQ5} reduces to
\begin{equation}
 \begin{aligned}
	\dot{c}_a(t)&=-\Gamma c_a(t)-\Gamma e^{i\theta_0}c_b(t-\tau_0 )\Theta (t-\tau_0 ),\\
	\dot{c}_b(t)&=-\Gamma c_b(t)-\Gamma e^{i\theta_0}c_a(t-\tau_0 )\Theta (t-\tau_0).\\
\end{aligned}
\label{EQ8}
\end{equation}

As seen from Eq.~\eqref{EQ5}, each giant atom typically has one self-interference term and four interatomic interference terms. Under the phase conditions of Case I, the self-interference terms vanish completely, and two of the four interatomic paths are suppressed. The system thus reduces to the effective dynamics of Eq.~\eqref{EQ8}. Figure~\figpanel{FIG2}{a} illustrates the remaining and suppressed paths.

Initializing the system in the symmetric state $|+\rangle = (|eg\rangle + |ge\rangle)/\sqrt{2}$. The concurrence $C_N^\mathrm{I}$ is plotted in Figs.~\figpanel{FIG3}{a} and~\figpanel{FIG3}{b} for various $\Gamma\tau_0$. In the Markovian limit ($\Gamma\tau_0=0$), the analytical solution is 
\begin{equation}
\begin{split}
C_N^\mathrm{I}(t)=e^{-2\Gamma(1+\cos\theta_0)t}.
\end{split}
\label{EQ9}
\end{equation}
As shown in Fig.~\figpanel{FIG3}{a} and Eq.~\eqref{EQ9}, significant entanglement storage occurs only near $\theta_0=\pi$. Compared to conventional nested giant atoms, the dependence on $\theta_0$ is smoother, with no prominent side peaks. 

In the non-Markovian regime ($\Gamma\tau_0=0.8$), $C_N^\mathrm{I}$ exhibits weak early-time oscillations and rapidly converges to a steady state at $\theta_0=\pi$. The steady-state value depends on the time delay:
\begin{equation}
\begin{split}
C_{N}^{\mathrm{I}}(\infty )=\frac{1}{(1+\Gamma \tau_0)^2}.
\end{split}
\label{EQ10}
\end{equation}
This behavior resembles that of conventional nested giant atoms ~\cite{PhysRevA.108.023728}, but with reduced oscillations and faster stabilization due to the suppression of certain paths. Thus, appropriate coupling phases mitigate non-Markovian effects and accelerate entanglement generation.

Notably, under the conditions of Case I, the entanglement dynamics become identical to those of two small atoms coupled to the waveguide with phase shift $\theta_0$ and time delay $\Gamma\tau_0$ [see Fig.~\figpanel{FIG2}{b}], except for an effective enhancement of the coupling strength from $g$ to $\sqrt{2}g$. This establishes a direct correspondence between the nested giant-atom system and a small-atom system.

In Case II ($\varphi_{a1}=\varphi_{b2}=0$ and $\varphi_{a2}=\varphi _{b1}=\pi/2$), the dynamical equations become
\begin{equation}
 \begin{aligned}
	\dot{c}_a(t)=-\Gamma c_a(t)-\Gamma e^{2i\theta_0}c_b(t-2\tau_0)\Theta (t-2\tau_0),\\
	\dot{c}_b(t)=-\Gamma c_b(t)-\Gamma e^{2i\theta_0}c_a(t-2\tau_0)\Theta (t-2\tau_0).\\
\end{aligned}
\label{EQ11}
\end{equation}
Here, the pattern of active and suppressed paths differs from Case I [compare Figs.~\figpanel{FIG2}{a} and~\figpanel{FIG2}{d}]. The concurrence dynamics are shown in Figs.~\figpanel{FIG3}{c} and~\figpanel{FIG3}{d}. In the Markovian limit $(\Gamma\tau_0=0)$, efficient entanglement storage occurs at $\theta_0=\pi/2$ and $\theta_0=3\pi/2$, with 
 \begin{equation}
\begin{split}
C_N^\mathrm{II}(t)=e^{-2\Gamma[1+\cos(2\theta_0)]t}.
\end{split}
\label{EQ12}
\end{equation}

In the non-Markovian effects ($\Gamma\tau_0=0.8$) regime, the steady-state value is 
\begin{equation}
\begin{split}
C_{N}^{\mathrm{II}}(\infty )=\frac{1}{(1+2\Gamma \tau_0)^2}.
\end{split}
\label{EQ13}
\end{equation}  
Again, the dynamics map onto those of two small atoms, now with an effective phase shift $2\theta_0$, as shown in Fig.~\figpanel{FIG2}{c}.

These two cases demonstrate that coupling phases enable selective control of propagation paths, reducing the impact of non-Markovian effects on entanglement.

\section{Robustness of entanglement against phase shifts}
\label{section4}

\subsection{Perfect Robustness in Nested Giant Atoms}

\begin{figure}[htbp]
    \centering
    \includegraphics[width=1\linewidth]{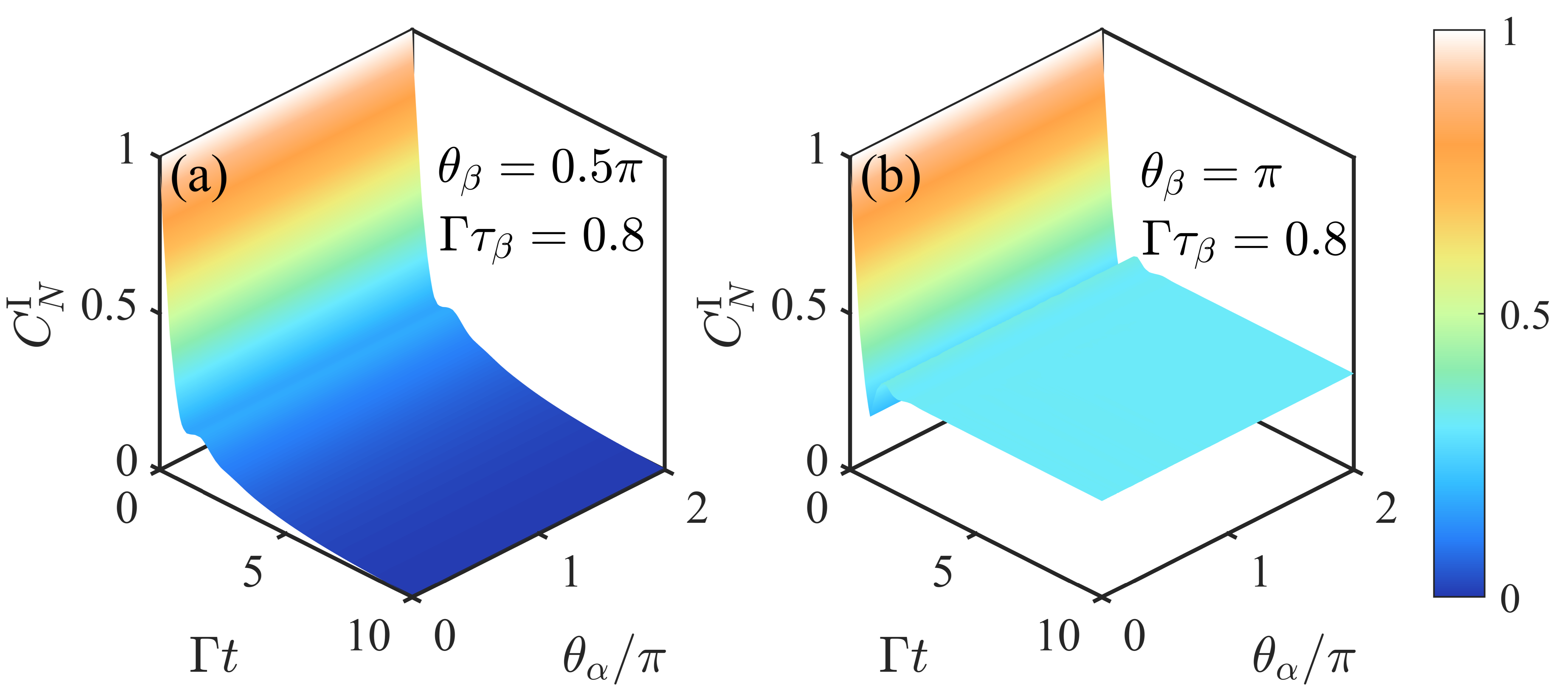}
    \caption{Concurrence $C_{N}^{\mathrm{I}}$ as a function of evolution time $\Gamma t$ and partial phase shift $\theta_\alpha$ for different $\theta_\beta$. Initial state $|+\rangle$ and time delay $\Gamma \tau_0=0.8$. The Time delay associated with $\theta_\alpha$ is $\tau_\alpha = \theta_\alpha \tau_\beta/\theta_\beta$. Coupling phases follow Case I.}
    \label{FIG4}
\end{figure}

We now consider the nested configuration of Fig.~\figpanel{FIG2}{a} with the phase conditions of Case I, but treat $\theta_{\alpha}$ and $\theta_{\beta}$ as independent. For fixed  $\theta_{\beta}$ and $\Gamma\tau_{\beta}$, the concurrence $C_{N}^{\mathrm{I}}$ is plotted as a function of $\theta_\alpha$ and $\Gamma t$ in Fig.~\ref{FIG4}.

The entanglement dynamics are completely insensitive to $\theta_\alpha$, demonstrating robustness against this parameter and  its associated time delay. As illustrated in Fig.~\figpanel{FIG2}{a}, only two interatomic paths survives, both with phase shift $\theta_\beta$ and time delay $\Gamma\tau_\beta$. Thus, contribution from paths involving $\theta_\beta$ are eliminated by destructive interference.

This effect is unique to the nested configuration. In separate or braided geometries, although four interatomic paths exist, at most two can be suppressed; the remaining two cannot be fully canceled. In the nested configuration, pairwise cancellation of paths is enabled by the intrinsic topology.

\subsection{Strong Robustness in Separate Giant Atoms}

\begin{figure}[htbp]
    \centering
    \includegraphics[width=1\linewidth]{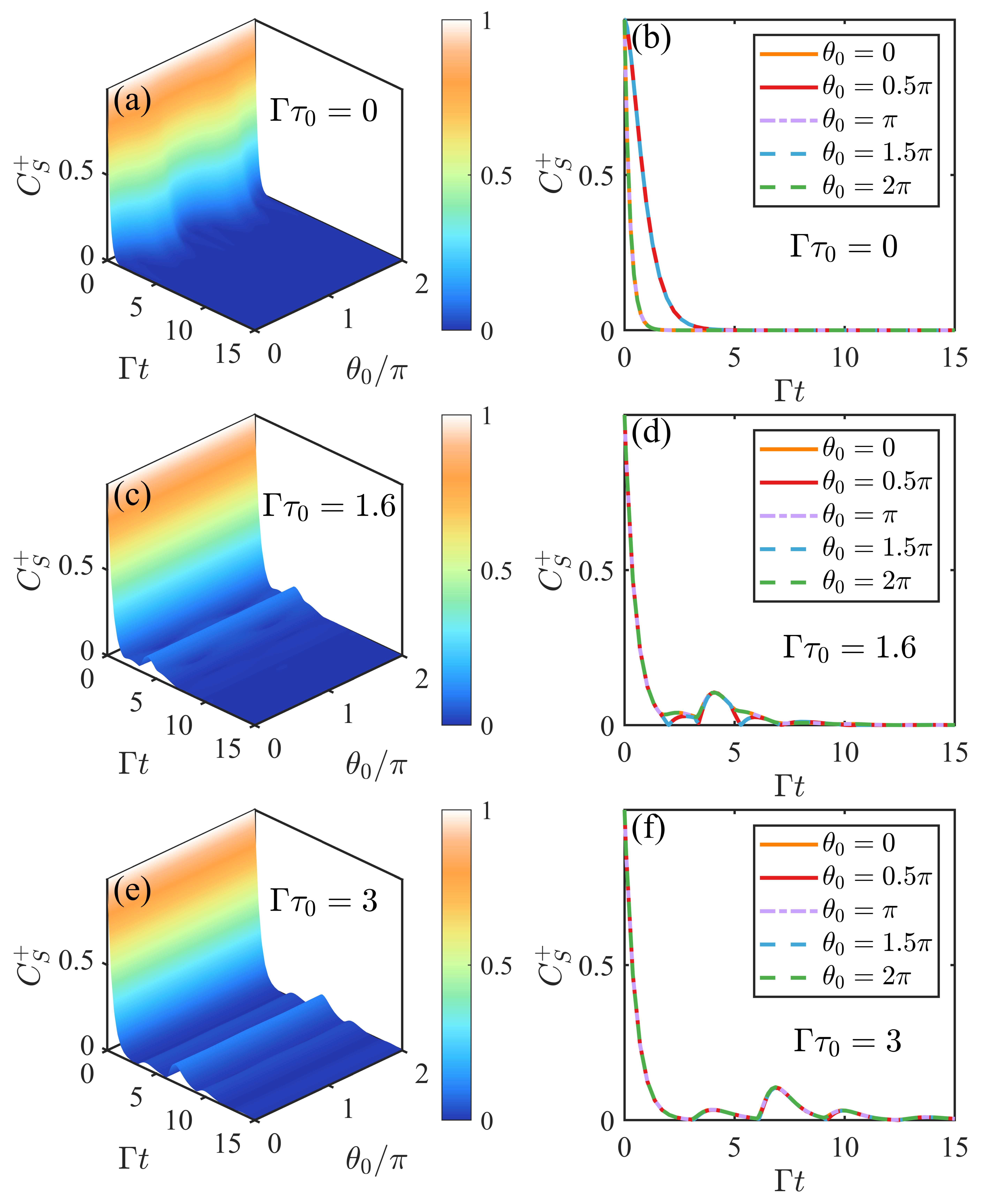}
    \caption{(a), (c), and (e) Concurrence $C_{S}^{+}$ as a function of the phase shift $\theta_0$ and the evolution time $\Gamma t$ for different time delays $\Gamma \tau_0$. (b), (d), and (f) Corresponding time traces. Parameters: initial state $|+\rangle$; coupling phases $\varphi_{a1}=\varphi_{b1}=0$ and $\varphi_{a2}=\varphi_{b2}=\pi/2$.}
    \label{FIG5}
\end{figure}

We now examine the separate configuration [see Fig.~\figpanel{FIG1}{a}] with coupling phases $\varphi_{a1}=\varphi_{b1}=0$, $\varphi_{a2}=\varphi _{b2}=\pi/2$.  Defining symmetric and antisymmetric amplitudes $\chi_{\pm} = (c_a \pm c_b)/\sqrt{2}$, Eq.~\eqref{EQ5} yields
\begin{equation}
\begin{split}
\begin{aligned}
	\dot{\chi}_+(t)=&-\Gamma \chi _+(t)+\frac{i\Gamma}{2}e^{i\theta _0}\chi _-(t-\tau _0)\Theta (t-\tau _0)\\ &-\Gamma e^{i2\theta _0}\chi _+(t-2\tau _0)\Theta (t-2\tau _0)\\ & -\frac{i\Gamma}{2}e^{i3\theta _0}\chi _-(t-3\tau _0)\Theta (t-3\tau _0),\\
	\dot{\chi}_-(t)=&-\Gamma \chi _-(t)-\frac{i\Gamma}{2}e^{i\theta _0}\chi _+(t-\tau _0)\Theta (t-\tau _0)\\
	& +\Gamma e^{i2\theta _0}\chi _-(t-2\tau _0)\Theta (t-2\tau _0)\\
	& +\frac{i\Gamma}{2}e^{i3\theta _0}\chi _+(t-3\tau _0)\Theta (t-3\tau _0).\\
\end{aligned}
\end{split}
\label{EQ14}
\end{equation}
The concurrence becomes  
\begin{equation}
\begin{split}
\begin{aligned}
C(t)=\left| |\chi _+(t)|^2-|\chi _-(t)|^2+2i\mathrm{Im[}\chi _+\left( t \right) \chi _-(t)^*] \right|.
\end{aligned}
\end{split}
\label{EQ15}
\end{equation}
Figure~\ref{FIG5} shows $C_{S}^{+}$ (concurrence for the symmetric initial state) as a function of $\theta_0$ and $\Gamma t$ for different $\Gamma\tau_0$. In the Markovian limit ($\Gamma\tau_0=0$), the delayed terms in Eq.~\eqref{EQ14} become instantaneous, yielding the simplified differential equations:
\begin{equation}
\begin{split}
\begin{aligned}
	\dot{\chi}_+(t)&=-\Gamma (1+e^{i2\theta _0})\chi _+(t)+\frac{i\Gamma}{2}\left( e^{i\theta _0}-e^{i3\theta _0} \right) \chi _-(t),\\
	\dot{\chi}_-(t)&=-\Gamma (1-e^{i2\theta _0})\chi _-(t)+\frac{i\Gamma}{2}\left( e^{i3\theta _0}-e^{i\theta _0} \right) \chi _+(t).
\end{aligned}
\end{split}
\label{EQ16}
\end{equation}
Substituting into Eq.~\eqref{EQ15} yields the evolution of $C_{S}^{+}$, as shown in Figs.~\figpanel{FIG5}{a} and~\figpanel{FIG5}{b}. The concurrence decays exponentially to zero over time for any phase shift $\theta_0$. The decay is fastest at $\theta_0=n\pi$ and slowest at $\theta_0=\pi/2 + n\pi$, where $n$ is an integer. The corresponding analytical expressions are
\begin{equation}
    C_{S}^{+}(t) = \begin{cases}
        \begin{aligned}
            & e^{-4\Gamma t}, && \theta_0=n\pi, \\
            & e^{-2\Gamma t} (2\Gamma t + 1), && \theta_0 =\frac{\pi}{2} + n\pi.
        \end{aligned}
    \end{cases}
    \label{eq17}
\end{equation}

When non-Markovian effects are included, the time-delay effect becomes more pronounced with increasing time delay, while the influence of $\theta_0$ on the $C_{S}^{+}$ gradually weakens, as shown in Figs.~\figpanel{FIG5}{c} and~\figpanel{FIG5}{e}. When the time delay reaches $\Gamma\tau_0=3$, the dynamics of $C_{S}^{+}$ becomes nearly independent of $\theta_0$, which is particularly evident in Fig.~\figpanel{FIG5}{f}. To clarify this behavior, we analyze the analytical solutions of $C_{S}^{+}$ dynamics in the interval $0\le \Gamma t\le3\Gamma\tau_0$.
 
In the interval $0<\Gamma t<\Gamma\tau_0$, the dynamical equations for the symmetric and antisymmetric amplitudes become
\begin{equation}
\begin{array}{l}
	\dot{\chi}_+(t)=-\Gamma \chi _+(t)\\
	\dot{\chi}_-(t)=-\Gamma \chi _-(t)\\
\end{array}.
\label{EQ18}
\end{equation}
In this case, the concurrence is $C_{S}^{+}(t)=e^{-2\Gamma t}$.  Owing to the time-delay effect, the two atoms are not yet connected through any propagation paths. Consequently, the phase shift $\theta_0$ has no influence on the evolution of the concurrence.

For $\Gamma\tau_0<\Gamma t<2\Gamma\tau_0$, the amplitudes satisfy
\begin{equation}
\begin{aligned}
\dot{\chi}_+(t) &= -\Gamma \chi_+(t) 
+ \frac{i\Gamma}{2} e^{i\theta_0}\chi_-(t-\tau_0)\Theta(t-\tau_0), \\
\dot{\chi}_-(t) &= -\Gamma \chi_-(t) 
- \frac{i\Gamma}{2} e^{i\theta_0}\chi_+(t-\tau_0)\Theta(t-\tau_0).
\end{aligned}
\label{EQ19}
\end{equation}
With initial conditions $\chi _+(\tau _0)=e^{-\Gamma \tau _0}$ and $\chi _-(\Gamma\tau _0)=0$, we obtain 
\begin{equation}
\begin{split}
C_{S}^{+}(t)=\Bigg\{
\left[e^{-2\Gamma t}-\frac{\Gamma ^2}{4}(t-\tau _0)^2
e^{-2\Gamma (t-\tau _0)}\right]^2 \\
+\left[\Gamma (t-\tau _0)e^{-\Gamma (2t-\tau _0)}
\cos\theta _0\right]^2
\Bigg\}^{1/2}.
\end{split}
\label{EQ20}
\end{equation}
In this interval, $\theta_0$ begins to influence the evolution. Notably, only the second term depends on $\theta_0$, and it contains an exponential decay factor $e^{-\Gamma (2t-\tau _0)}$. As $\Gamma\tau_0$ increases, this factor decreases, significantly suppressing the influence of $\theta_0$ on the concurrence dynamics.

For $2\Gamma\tau_0<\Gamma t<3\Gamma\tau_0$, with initial conditions $\chi_+(2\tau_0)=e^{-2\Gamma\tau_0}$ and $\chi_-(2\tau_0)=-i\frac{\Gamma}{2}e^{i\theta_0}\tau_0 e^{-\Gamma\tau_0}$, we have
\begin{equation}
\begin{aligned}
\chi_+(t) &= e^{-\Gamma t} \bigg\{ 1 + e^{2\Gamma \tau_0} \big[ 
-\Gamma e^{i2\theta_0}(t - 2\tau_0) \\
&\quad + \tfrac{\Gamma^2}{8} e^{i2\theta_0} (t - 2\tau_0)^2 \big] \bigg\}, \\
\chi_-(t) &= -\tfrac{i\Gamma}{2}(t - \tau_0) e^{i\theta_0} e^{-\Gamma(t - \tau_0)},
\end{aligned}
\label{EQ21}
\end{equation}
Substituting into Eq.~\eqref{EQ15} gives the analytical expression for $C_{S}^{+}$. All terms containing $\theta_0$ are accompanied by strong exponential decay factors. As $\Gamma\tau_0$ increases, these terms are progressively suppressed, reducing the influence of $\theta_0$ on the entanglement dynamics.

 The three right panels of Fig.~\ref{FIG5} further support this analysis. In particular, the coupling phase $\varphi_{jn}$ can effectively reduce the influence of the phase shift $\theta_0$ on the concurrence dynamics. Moreover, in the presence of appropriate non-Markovian effects, the robustness of the concurrence evolution against $\theta_0$ is further enhanced.

\section{Equivalent entanglement dynamics in different giant-atom configurations induced by phase modulation}
\label{section5}

\begin{figure}[t]
    \centering
    \includegraphics[width=0.85\linewidth]{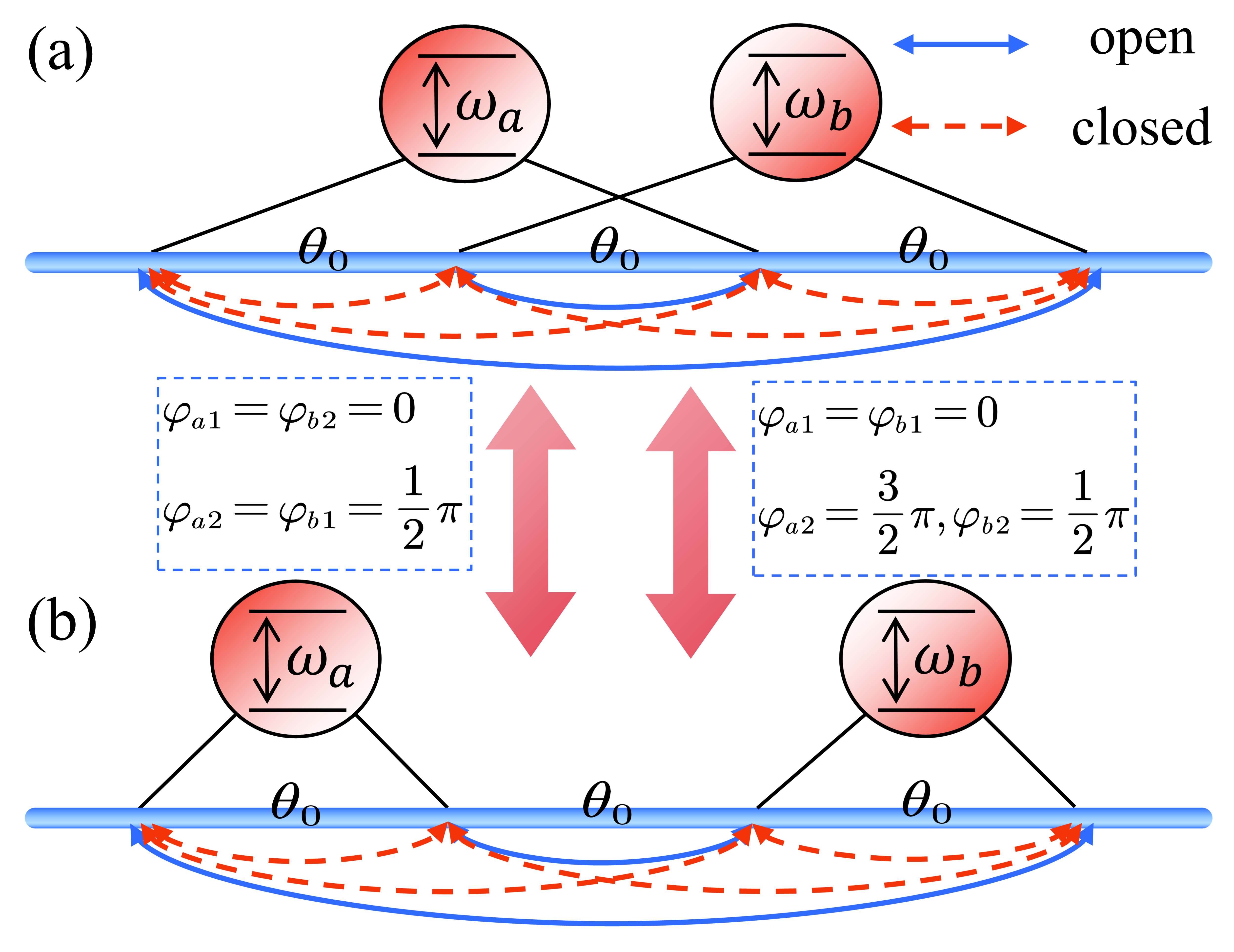}
    \caption{ Separate and braided configurations contain both common and distinct propagation paths. By choosing appropriate coupling phases, the distinct paths can be suppressed via destructive interference, leaving only the common ones and yielding identical entanglement dynamics.}
    \label{FIG6}
\end{figure}

As shown in Sec.~\ref{section2}, the coupling phase significantly affects the propagation paths between atoms. Here we investigate both the separate and braided giant-atom configurations. For simplicity, the phase shift between any two coupling points is taken as $\theta_0$, and the time delay as $\Gamma\tau_0$. When Case I is chosen, i.e., $\varphi _{a1}=\varphi _{b2}=0$ and $\varphi _{a2}=\varphi _{b1}=\pi/2$, the concurrence dynamics of the two giant atoms become identical in the two configurations. In this case, the dynamical equations satisfy
\begin{equation}
\begin{aligned}
\dot{c}_a(t) ={}& -\Gamma c_a(t)
-\frac{\Gamma}{2}e^{i\theta _0}c_b(t-\tau _0)\Theta (t-\tau _0) \\
&-\frac{\Gamma}{2}e^{3i\theta _0}c_b(t-3\tau _0)\Theta (t-3\tau _0), \\
\dot{c}_b(t) ={}& -\Gamma c_b(t)
-\frac{\Gamma}{2}e^{i\theta _0}c_a(t-\tau _0)\Theta (t-\tau _0) \\
&-\frac{\Gamma}{2}e^{3i\theta _0}c_a(t-3\tau _0)\Theta (t-3\tau _0).
\end{aligned}
\label{EQ22}
\end{equation}
This phenomenon can be understood from Fig.~\ref{FIG6}. Although the two configurations contain the same number of propagation paths, most are different and only a few are identical. Because the opening and closing of each propagation path are solely controlled by the coupling phase, an appropriate choice of the coupling phases can suppress the different paths through destructive interference, leaving only the identical ones. This is also confirmed by Eq.~\eqref{EQ15}. Consequently, the atomic dynamical evolution in the two configurations becomes identical, resulting in the same concurrence dynamics.

\begin{figure}[t]
    \centering
    \includegraphics[width=1\linewidth]{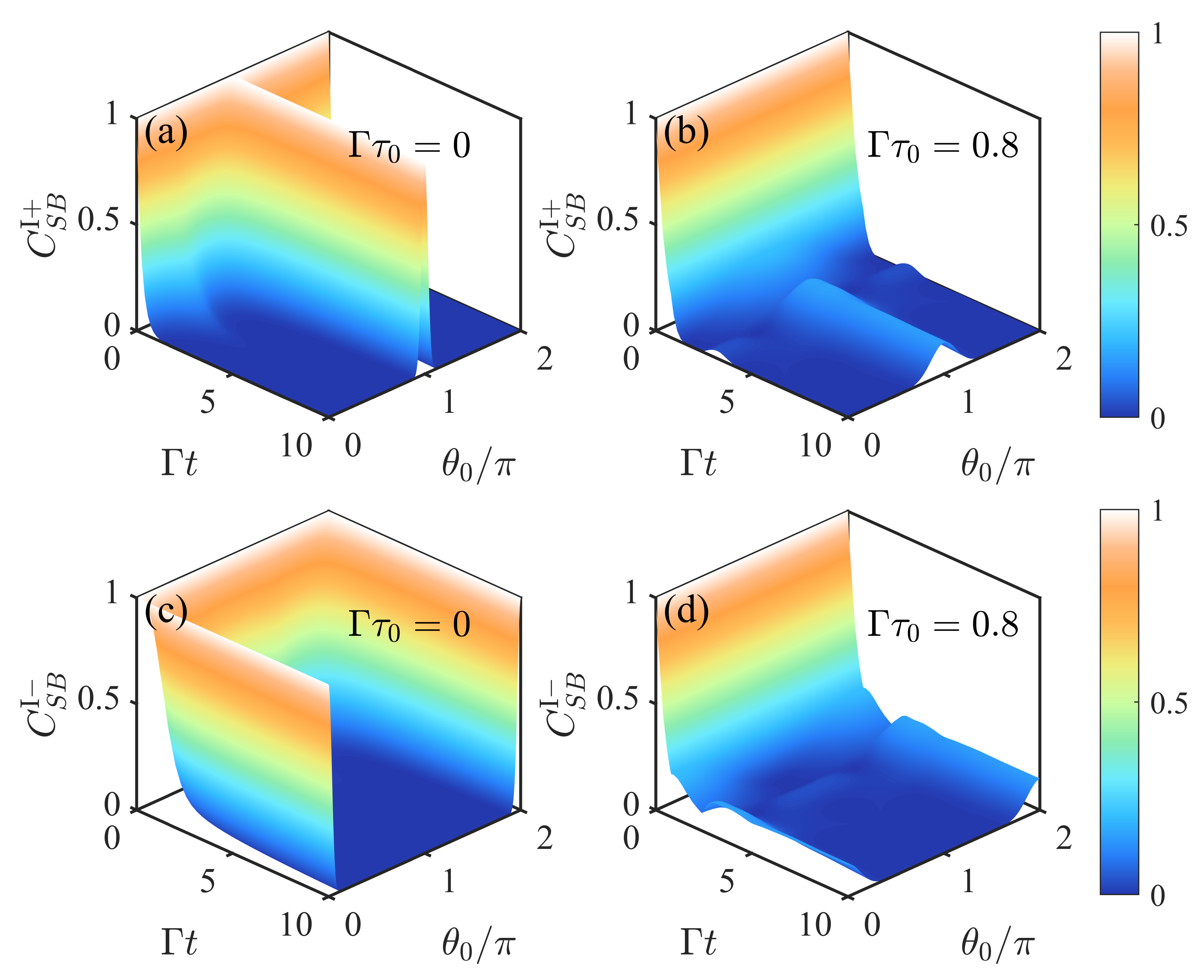}
    \caption{Concurrence $C_{SB}^\mathrm{I\pm}$ as a function of the evolution time $\Gamma t$ and phase shift $\theta_0$ for different initial states and time delay $\Gamma\tau_0$. (a) and (b): symmetric initial state $|+\rangle$; (c) and (d): antisymmetric initial state $|-\rangle$. (a) and (c): $\Gamma\tau_0=0$; (b) and (d): $\Gamma\tau_0=0.8$. Case I: $\varphi_{a1}=\varphi_{b2}=0$ and $\varphi_{a2}=\varphi_{b1}=\pi/2$.}

    \label{FIG7}
\end{figure}

In the Markovian limit ($\Gamma\tau_0=0$), with the symmetric initial state $|+\rangle$ for both configurations, the concurrence is
\begin{equation}
\begin{array}{c}
	C_{SB}^{\mathrm{I}+}(t)=e^{-2\Gamma \left[ 1+\frac{1}{2}\cos \theta _0+\frac{1}{2}\cos \left( 3\theta _0 \right) \right] t}.
\end{array}
\label{EQ23}
\end{equation}
From Eq.~\eqref{EQ23} and Fig.~\figpanel{FIG7}{a}, efficient entanglement storage occurs only at $\theta_0=\pi$. When non-Markovian effects are included, $C_{SB}^\mathrm{I+}$ still exhibits pronounced entanglement storage at the same phase shift after a short transient oscillation, indicating that the coupling phase suppresses the influence of non-Markovian effects. The steady-state value is
\begin{equation}
\begin{split}
\begin{aligned}
C_{SB}^\mathrm{I+}(t\rightarrow \infty )=\frac{1}{(1+2\Gamma \tau _0)^2}.
\end{aligned}
\end{split}
\label{EQ24}
\end{equation}

For the antisymmetric initial state $|-\rangle$, the concurrence dynamics exhibit markedly different behavior. In the Markovian limit, efficient storage occurs only at $\theta_0=2\pi$, with
\begin{equation}
\begin{array}{c}
C_{SB}^{\mathrm{I}-}(t)=e^{-2\Gamma \left[ 1-\frac{1}{2}\cos \theta _0-\frac{1}{2}\cos \left( 3\theta _0 \right) \right] t}.
\end{array}
\label{EQ25}
\end{equation}
When non-Markovian effects are taken into account, the steady-state value of the stored concurrence is the same as that for the symmetric initial state.

\begin{figure}[t]
    \centering
    \includegraphics[width=1\linewidth]{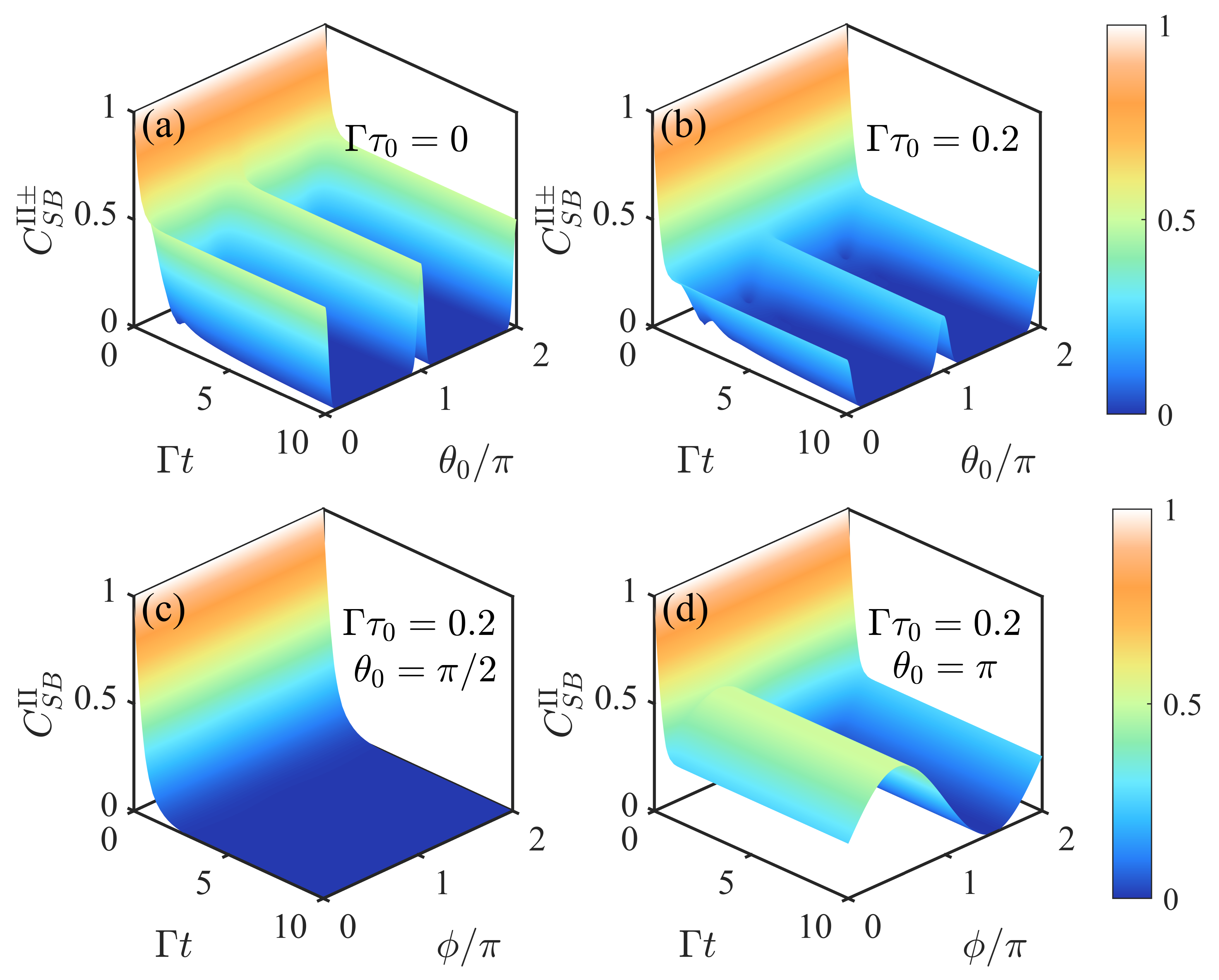}
    \caption{(a) and (b) Concurrence $C_{NB}^\mathrm{II\pm}$ as a function of the evolution time $\Gamma t$ and the phase shift $\theta_0$ for symmetric or antisymmetric state. (c) and (d) Concurrence $C_{NB}^{\mathrm{II}}$ versus $\Gamma t$ and $\phi$ for different $\theta_0$, with initial state $(|eg\rangle + e^{i\phi}|ge\rangle)/\sqrt{2}$. $\Gamma t=0$ in (a) and $\Gamma t=0.2$ in (b)–(d). In (c) and (d), $\theta_0=\pi/2$ and $\pi$, respectively. Case II: $\varphi_{a1}=\varphi_{b1}=0$, $\varphi_{a2}=3\pi/2$, $\varphi_{b2}=\pi/2$.}
    \label{FIG8}
\end{figure}

In Case II, where the additional phases are chosen as $\varphi_{a1}=\varphi _{b1}=0$, $\varphi_{a2}=3\pi/2$, and $\varphi_{b2}=3\pi/2$, the dynamical equations of the two configurations remain completely equivalent, resulting in identical $C_{SB}^\mathrm{II\pm}$ dynamics. In this case, the dynamical equations satisfy
\begin{equation}
\begin{aligned}
\dot{c}_a(t)={}&-\Gamma c_a(t)+i\frac{\Gamma}{2}e^{i\theta _0}c_b(t-\tau _0)\Theta (t-\tau_0)\\&+i\frac{\Gamma}{2}e^{3i\theta _0}c_b(t-3\tau _0)\Theta (t-3\tau _0), \\
\dot{c}_b(t)={}&-\varGamma c_b(t)-i\frac{\Gamma}{2}e^{i\theta _0}c_a(t-\tau _0)\Theta (t-\tau _0)\\&-i\frac{\Gamma}{2}e^{3i\theta _0}c_a(t-3\tau _0)\Theta (t-3\tau _0).
\end{aligned}
\label{EQ26}
\end{equation}
The mechanism responsible for the identical entanglement dynamics is the same as in Case I, as illustrated in Fig.~\ref{FIG6}. Interestingly, $C_{NB}^\mathrm{II\pm}$ dynamics remain identical for both the symmetric and antisymmetric initial states. In the Markovian limit ($\Gamma\tau_0=0$), $C_{NB}^\mathrm{II\pm}$ is given by
\begin{equation}
\begin{aligned}
C_{SB}^\mathrm{II\pm}(t) = e^{-2\Gamma t}\sqrt{\cos ^2\left[ 2\mathrm{Im}\left( z \right) \right] +\sinh ^2\left[ 2\mathrm{Re}\left( z \right) \right]},
\end{aligned}
\end{equation}
where $z=\Gamma te^{2i\theta_0} \cos \theta_0 $.  In this case, efficient storage occurs at $\theta_0=\pi+n\pi$, as shown in Fig.~\figpanel{FIG8}{a}. When non-Markovian effects are included, $C_{SB}^\mathrm{II\pm}$ exhibits almost no oscillations at the early stage of the evolution due to the coupling phase [see Fig.~\figpanel{FIG8}{b})]. However, the steady-state value is still influenced by non-Markovian effects and is given by
\begin{equation}
\begin{split}
\begin{aligned}
C_{SB}^\mathrm{II\pm}(t\rightarrow \infty )=\frac{1}{2(1+2\Gamma \tau _0)^2}.
\end{aligned}
\end{split}
\label{EQ28}
\end{equation}

We further consider the general entangled initial state 
$(|eg\rangle+e^{i\phi}|ge\rangle)/\sqrt{2}$ and study the influence of the phase $\phi$ on the concurrence dynamics in the presence of non-Markovian effects. For $\theta_0=\pi/2$, the phase $\phi$ has almost no effect on the concurrence dynamics, and $C_{SB}^\mathrm{II}$ rapidly decays to zero over time [Fig.~\figpanel{FIG8}{c}], consistent with the behavior shown in Fig.~\figpanel{FIG8}{a}. When $\theta_0=\pi$, however, $\phi$ significantly affects the steady-state value [Fig.~\figpanel{FIG8}{d}], which is given by
\begin{equation}
\begin{split}
\begin{aligned}
C_{SB}^\mathrm{II}(t\rightarrow \infty )=\frac{1+\sin \phi}{2(1+2\Gamma \tau _0)^2}
\end{aligned}
\end{split}
\label{EQ29}
\end{equation}

Both schemes can realize equivalent concurrence dynamics between the separate and braided giant-atom configurations. Although the same propagation paths are preserved in the two configurations, the corresponding coupling phases differ, as seen from Eqs.~\eqref{EQ22} and~\eqref{EQ26}. Consequently, the resulting entanglement dynamics still exhibit certain differences.

\section{DISCUSSION AND CONCLUSION}
\label{section6}

We have investigated three typical configurations of two giant atoms coupled to a waveguide, with additional coupling phases introduced at each coupling point. Unlike previous studies on chiral giant atoms, which mainly focus on nonreciprocal transport or chiral spontaneous emission, this work emphasizes the role of coupling phases in controlling the propagation paths between atoms. Specifically, each propagation path is jointly affected by two additional phases, enabling destructive interference between otherwise independent paths under appropriate conditions. This interference mechanism suppresses selected propagation paths and thereby weakens the influence of non-Markovian effects in the system. As a consequence, the oscillations of entanglement at the early stage of evolution are significantly weakened, and the system approaches its steady state more rapidly.

Furthermore, the effects of coupling-phase modulation exhibit distinct features in different configurations. In the nested configuration, appropriate phase tuning can eliminate most propagation paths, making the entanglement dynamics equivalent to those of two small atoms. This originates from the suppression of contributions from certain waveguide modes, rendering the system dynamics robust against the phase shift parameter $\theta_{\beta}$. In the separated configuration, the interplay between coupling phases and non-Markovian effects similarly enhances the robustness of entanglement against phase shift variations. Moreover, by properly engineering the coupling phases to suppress specific propagation paths, identical entanglement dynamics can be achieved between the braided and separated configurations. This work offers a new perspective for the experimental generation of stable entanglement and the design of quantum devices with programmable propagation and controllable memory effects.

\section*{Acknowledgments}

We thank Dr. Lei Du for fruitful discussions. This work is supported by the National Natural Science Foundation of China  (Grants No. 12564047, No. 11874004, No. 11204019, and No. 12564048).

\bibliography{GA_ref.bib}
\end{document}